\documentstyle[epsfig]{aipproc2}

\input epsf

\ifx\epsffile\undefined
\newlength{\epsfysize}
\def\epsffile#1#2#3#4]#5{}
\fi

\begin{document}

\newcommand{\be}{\begin{equation}}
\newcommand{\ee}{\end{equation}}
\newcommand{\bear}{\begin{eqnarray}}
\newcommand{\eear}{\end{eqnarray}}
\newcommand{\gsim}{\lower.7ex\hbox{$\;\stackrel{\textstyle>}{\sim}\;$}}
\newcommand{\lsim}{\lower.7ex\hbox{$\;\stackrel{\textstyle<}{\sim}\;$}}
\newcommand{\dr}{\mbox{\footnotesize{$\overline{\rm DR}$\ }}}
\newcommand{\GeV}{{\rm GeV}}
\newcommand{\TeV}{{\rm TeV}}
\newcommand{\met}{\not\!\!\!E_T}
\newcommand{\msusy}{M_{\rm SUSY}}

\def\cpc#1 #2 #3 #4 {Comp.~Phys.~Comm.  {\bf  #1}, #2 (#3)#4 }
\def\npb#1 #2 #3 #4 {Nucl.~Phys. {\bf B#1}, #2 (#3)#4 }
\def\plb#1 #2 #3 #4 {Phys.~Lett. {\bf B#1}, #2 (#3)#4 }
\def\prd#1 #2 #3 #4 {Phys.~Rev.  {\bf D#1}, #2 (#3)#4 }
\def\prl#1 #2 #3 #4 {Phys.~Rev.~Lett. {\bf #1}, #2 (#3)#4 }
\def\pr#1  #2 #3 #4 {Phys.~Rept. {\bf #1}, #2 (#3)#4 }
\def\mpl#1 #2 #3 #4 {Mod.~Phys.~Lett. {\bf A#1}, #2 (#3)#4 }
\def\zpc#1 #2 #3 #4 {Z.~Phys. {\bf C#1}, #2 (#3)#4 }
%

\rightline{FERMILAB-PUB-00/003-T}  
\rightline{January 20, 2000}    
\rightline{hep-ex/0001007}  

\title{Discovering a Light Higgs Boson with Light}

\author{Greg Landsberg$^1$ and Konstantin T.~Matchev$^2$}
\address{$^1$Physics Department,
             Brown University, 182 Hope St, Providence, RI 02912\\
         $^2$Theoretical Physics Department,
             Fermi National Accelerator Laboratory, Batavia, IL 60510}

\maketitle

\begin{abstract}
We evaluate the prospects for detecting a non-standard light
Higgs boson with a significant branching ratio to two photons,
in the Run II of the Fermilab Tevatron. We derive the reach
for several channels: $2\gamma$ inclusive, $2\gamma+1$ jet,
and $2\gamma+2$ jets. We present the expected Run II limits
on the branching ratio of $h\rightarrow\gamma\gamma$
as a function of the Higgs mass, for the case of ``bosonic'',
as well as ``topcolor'' Higgs bosons.
\end{abstract}

\section*{Motivation}

The Standard Model (SM) is very economical in the sense that
the Higgs doublet responsible for electroweak symmetry breaking
can also be used to generate fermion masses.
The Higgs boson couplings to the gauge bosons,
quarks, and leptons are therefore predicted in the
Standard Model, where one expects the Higgs boson
to decay mostly to b-jets and tau pairs (for low Higgs
masses, $M_h\lsim 140$ GeV), or to $WW$ or $ZZ$ pairs,
(for higher Higgs masses, $M_h \gsim 140$ GeV).
Since the Higgs boson is neutral and does not couple to
photons at tree level, the branching ratio
${\rm B}(h\rightarrow \gamma\gamma)$ is predicted to be
very small in the SM, on the order of $10^{-3}-10^{-4}$.

In a more general framework, however,
where different sectors of the theory are responsible
for the physics of flavor and electroweak symmetry breaking,
one may expect deviations from the SM predictions, which may
lead to drastic changes in the Higgs boson discovery signatures.
One such example is the so called ``fermiophobic''
(also known as ``bosophilic'' or ``bosonic'') Higgs,
which has suppressed couplings to all fermions.
It may arise in a variety of models, see e.g. 
\cite{bosmodels}.
A variation on this theme is the Higgs in certain
topcolor models, which may couple to heavy quarks only
\cite{topmodels}. Some even more exotic possibilities
have been suggested in the context of theories with
large extra dimensions \cite{LED}.
Finally, in the minimal supersymmetric standard model (MSSM),
the width into $b\bar{b}$ pairs can be suppressed
due to 1-loop SUSY corrections, thus enhancing the
branching ratios of a light Higgs into more exotic signatures
\cite{CMW,Mrenna}. 
In all these cases, the Higgs boson decays to photon pairs
are mediated through a $W$ or heavy quark loop and dominate for
$M_h\lsim 100$ GeV \cite{SMW}. In the range $100\lsim M_h\lsim 160$,
they compete with the $WW^\ast$ mode, while
for $M_h\gsim 160$ GeV, $h\rightarrow WW$ completely takes over.
Current bounds from LEP \cite{LEP limits} are limited by the
kinematic reach of the machine. The existing Run I analyses 
at the Tevatron have utilized the diphoton plus
2 jets \cite{Lauer,D0,Wilson} and inclusive diphoton \cite{Wilson}
channels and were limited by statistics.
Since they only looked for a ``bosonic'' Higgs
\cite{bosmodels}, they did not consider the Higgs production
mechanism through gluon fusion, which can be a major additional
source of signal in certain models \cite{topmodels}.
Since $h\rightarrow \gamma\gamma$ is a very clean signature,
it will allow the Tevatron to extend
significantly those limits in its next run.

In this study we shall evaluate the Higgs discovery potential
of the upcoming Tevatron runs for several diphoton channels.
We shall concentrate on the following two questions. First,
what is the absolute reach in Higgs mass as a function of
the $h\rightarrow\gamma\gamma$ branching ratio? Second,
which signature (inclusive diphotons, diphotons plus one jet,
or diphotons plus two jets) provides the best reach.
We believe that neither of those two questions has been 
adequately addressed in the literature previously.

\section*{Tevatron Reach for a Bosonic Higgs}

Here we consider the case of a ``bosonic'' Higgs, 
i.e. models where the Higgs couplings to all fermions are 
suppressed. Then, the main Higgs production modes
at the Tevatron are associated $Wh/Zh$
production, as well as $WW/ZZ$ fusion.
All of these processes have comparable rates
\cite{Spira}, so it makes sense to consider
an inclusive signature first \cite{Wilson}.

\subsection*{Inclusive channel: analysis cuts}

We use the following cuts for our inclusive study:
two photons with $p_T(\gamma)>20$ GeV and rapidity
$|\eta(\gamma)| < 2$, motivated by the acceptance
of the CDF or D\O\ detectors in Run II. Triggering on such
a signature is trivial; both collaborations will have
diphoton triggers that are nearly fully efficient
with such offline cuts.

We assume 80\% diphoton identification efficiency,
which we apply to both the signal and background
estimates on top of the kinematic and geometrical acceptance.
Again, this efficiency is motivated by the CDF/D\O\ EM ID
efficiency in Run I and is not likely to change in Run II.

\subsection*{Inclusive channel: background}

The main backgrounds to the inclusive diphoton channel come
from the QCD production of dijets, direct photons, and diphotons.
In the former two cases a jet mimics a photon by fragmenting
into a leading $\pi^0/\eta$ meson that further decays into
a pair of photons, not resolved in the calorimeter.
\begin{figure}[t]
\epsfysize=3.5in
\epsffile[-250 0 100 560]{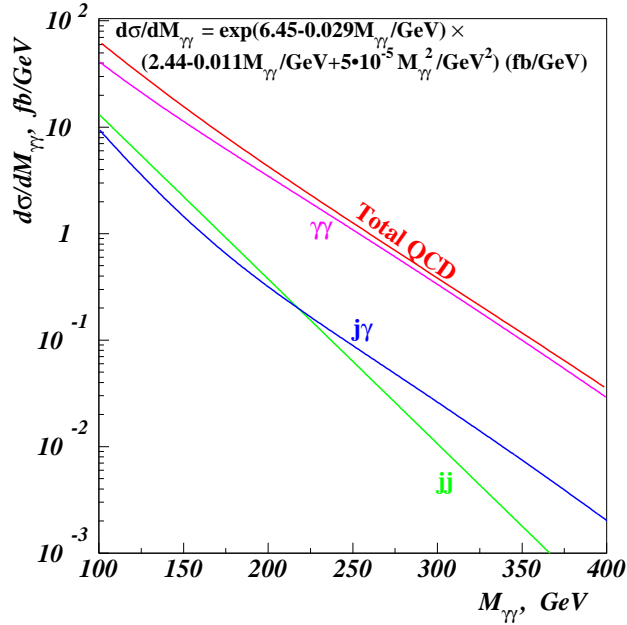}
\begin{center}
\caption[]
{\small The total background in the inclusive diphoton channel,
as well as the individual contributions from $\gamma\gamma$,
$\gamma j$ and $jj$ production. \label{qcd_background}}
\end{center}
\end{figure}

We used the PYTHIA~\cite{PYTHIA} event generator and the experimentally
measured probability of a jet to fake a photon~\cite{Lauer} to calculate
all three components of the QCD background. The faking probability depends
significantly on the particular photon ID cuts, especially on the 
photon isolation requirement~(see, e.g. \cite{Lauer,diboson,monopole}).
For this study we used an $E_T$-dependent jet-faking-photon probability of
$$
P({\rm jet} \rightarrow \gamma) =
\exp\left(-0.01\ {E_T\over\mbox{(1 GeV)}}-7.5\right),
$$
which is obtained by taking the $\eta$-averaged faking probabilities used in
the D\O\ Run I searches~\cite{Lauer}. The 
fractional error on $P(\mbox{jet} \rightarrow \gamma)$ is
about 25\% and is dominated by the uncertainty on the direct photon fraction
in the $\mbox{jet}+\gamma$ sample used for its determination.
(For high photon $E_T$,
however, the error is dominated by the available statistics.)
This probability is expected to remain approximately the same
in Run II for both the CDF and D\O\ detectors.
We used 80\% ID efficiency for the pair of photons,
and required the photons to
be isolated from possible extra jets in the event.
We accounted for NLO corrections via a constant $k$-factor of 1.34.

\begin{figure}[t]
\epsfysize=3.5in
\epsffile[-250 0 100 560]{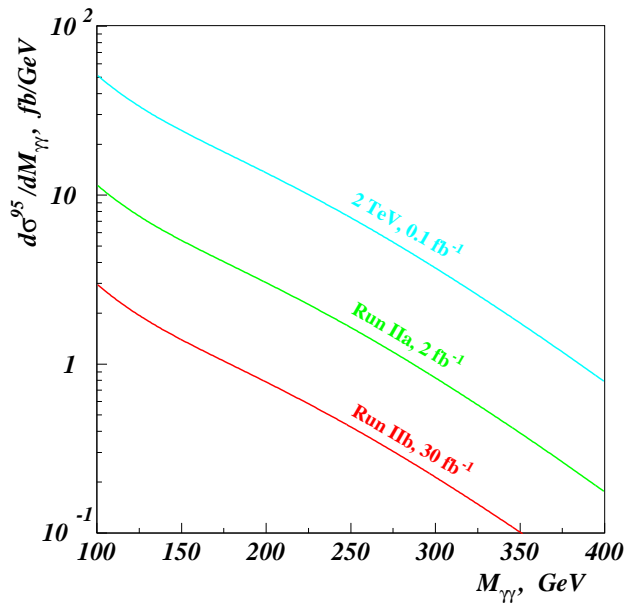}
\begin{center}
\caption[]
{\small The 95\% CL upper limit on $\varepsilon\times\sigma(\gamma\gamma+X)$
as a function of $M_{\gamma\gamma}$, for several benchmark
total integrated luminosities in Run II. 
\label{ggX}}
\end{center}
\end{figure}

Adding all background contributions, for the total background in the inclusive
diphoton channel we obtain the following parametrization:
$$
{d\sigma\over dM_{\gamma\gamma}}
= \biggl[p_3+p_4 \left({M_{\gamma\gamma}\over 1\ {\rm GeV}}\right)
      +p_5 \left({M_{\gamma\gamma}\over 1\ {\rm GeV}}\right)^2\biggr]\
\exp\,\biggl\{ p_1+
    p_2 \left( {M_{\gamma\gamma}\over 1\ {\rm GeV}}\right)\biggr\},
$$
where $p_1= 6.45$, $p_2=-0.029$, $p_3= 2.44$, $p_4=-0.011$ and $p_5= 0.00005$.
In the region $M_{\gamma\gamma}>100$ GeV it is dominated by direct
diphoton production and hence is irreducible.
The expected statistical plus systematic error on this background determination
is at the level of 25\%, based on the jet-faking photon probability uncertainty.
For larger invariant masses, however, the accuracy is dominated by the uncertainties
in the direct diphoton production cross section, which will be difficult to measure
independently in Run II, so one will still have to rely on the NLO predictions.
On the other hand, for narrow resonance searches one could do self-calibration of
the background by calculating the expected background under the signal peak via
interpolation of the measured diphoton mass spectrum between the regions just below
and just above the assumed resonance mass. Therefore, in our case the background
error
will be purely dominated by the background statistics. A combination of the
interpolation tecnique and the shape information from the theoretical NLO calculations of
the direct diphoton cross section is expected to result in significantly smaller
background error in Run II.

The total background, as well as the individual contributions from
$\gamma\gamma$, $\gamma j$ and $jj$ production, are shown
in Fig.~\ref{qcd_background}.
Additional SM background sources to the inclusive diphoton channel
include Drell-Yan production with both electrons misidentified
as photons, $W\gamma\gamma$ production, etc. and are all
negligible compared to the QCD background.
The absolute normalization of the background obtained by the
above method agrees well with the actual background
measured by CDF and D\O\ in the diphoton
mode~\cite{Wilson,monopole}.

In Fig.~\ref{ggX} we show the 95\% CL upper limit on the
differential cross section after cuts
$d(\varepsilon\times\sigma(\gamma\gamma+X))/dM_{\gamma\gamma}$
as a function of the diphoton invariant mass $M_{\gamma\gamma}$,
given the above background prediction (here $\varepsilon$
is the product of the acceptance and all efficiencies).
This limit represents $1.96\sigma$ sensitivity to a narrow signal
when doing a counting experiment in 1 GeV diphoton mass bins.
This plot can be used to obtain the sensitivity to any resonance
decaying into two photons as follows. One first fixes the width
of the mass window around the signal peak which is used in the analysis.
Then one takes the average value of the 95\% C.L. limit in
$d\sigma/dM_{\gamma\gamma}$ across the mass window from Fig.~\ref{ggX}
and multiplies it by $\sqrt{w/\mbox{GeV}}$, where $w$ is
the width of the mass window\footnote{The square root enters
the calculation since the significance is proportional to
the background to the $-1/2$ power.},
to obtain the corresponding 95\% CL
upper limit on the signal cross-section after cuts.
Similar scaling could be used if one is interested
in the 3$\sigma$ or 5$\sigma$ reach.

\subsection*{What is the optimum mass window cut?}

\begin{figure}[t]
\epsfysize=2.5in
\epsffile[-150 205 200 565]{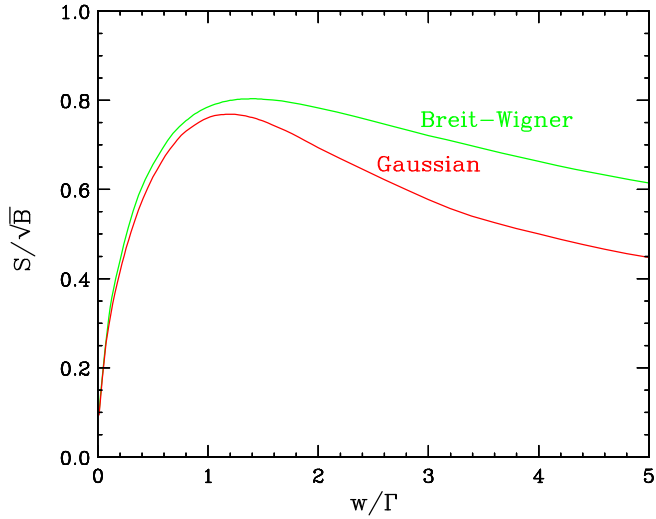}
\begin{center}
\caption[]
{\small Significance $S/\sqrt{B}$ (in arbitrary units),
as a function of the mass window width $w$ (in units of $\Gamma$),
for a Breit-Wigner or Gaussian resonance. \label{fig:significance}}
\end{center}
\end{figure}
When searching for narrow resonances in the presence of large
backgrounds ($B$), the best sensitivity toward signal ($S$)
is achieved by performing an unbinned maximum likelihood
fit to the sum of the expected signal and background shapes.
However, simple counting experiments give similar sensitivity
if the size of the signal ``window'' is optimized. For narrow
resonances the observed width\footnote{Notice that the width is
defined so that the cross-section at $\pm\Gamma/2$ away from the peak
is a factor of 2 smaller than the peak value (FWHM). For a Gaussian
resonance the width is related to the variance $\sigma$
by $\Gamma=2\sigma\sqrt{\ln4}\simeq 2.35\sigma$.}
$\Gamma$ is dominated by the instrumental effects, and is often Gaussian.
The background in a narrow window centered on the assumed
position $M_0$ of the peak in the signal invariant mass
distribution could be treated as linear. Therefore,
the Gaussian significance of the signal, $S/\sqrt{B}$,
as a function of the window width, $w$, is given by:
\begin{equation}
{S\over \sqrt{B}} 
\sim {1\over\sqrt{w}}\ 
{1\over \sqrt{2\pi}\sigma}
\int_{M_0-w/2}^{M_0+w/2}d\sqrt{s} 
\exp\left(-{(\sqrt{s}-M_0)^2\over 2\sigma^2}\right)
\sim {1\over\sqrt{w/\Gamma}}\ 
{\rm erf}\left(\sqrt{\ln2}\ {w\over\Gamma}\right),
\label{signgaus}
\end{equation}
where erf$(x)$ is the error function
$$
{\rm erf}(x)={2\over\sqrt{\pi}}\int_0^x e^{-t^2} dt.
$$
The function (\ref{signgaus}) is shown in
Fig.~\ref{fig:significance}, and has a maximum at $w \approx 1.2\Gamma$,
which corresponds to $\pm 1.2(\Gamma/2)$ cut around the resonance maximum.

For resonances significantly wider than the experimental resoluton,
the shape is given by the Breit-Wigner function, and in this case
the significance is:
\begin{equation}
{S\over\sqrt{B}}
\sim {1\over \sqrt{w}}
\int_{(M_0-w/2)^2}^{(M_0+w/2)^2}
{ds\over (s-M_0^2)^2+M_0^2\Gamma^2}
\sim {1\over \sqrt{w/\Gamma}}\arctan({w\over\Gamma}).
\end{equation}
This function, also shown in Fig.~\ref{fig:significance},
peaks at a similar value of $w$ ($w \approx 1.4\Gamma$).
We see that for both Gaussian and Breit-Wigner resonances,
the significance does not appreciably change when
using a $w = 1\Gamma-2\Gamma$ cuts. For our analysis
we shall use two representative choices: $w=1.2\Gamma$ 
and $w= 2\Gamma$ for the mass window, which we shall
always center on the actual Higgs mass. 

Clearly, one can do even better in principle, by suitably resizing and
repositioning the mass window around the bump in the combined $S+B$
distribution. Because of the steeply falling parton luminosities,
the signal mass peak is skewed and its maximum will appear somewhat
below the actual physical mass. In our analysis we choose
not to take advantage of these slight improvements,
thus accounting for unknown systematics.

\subsection*{Inclusive channel: results}

In Tables \ref{1sigma} and \ref{2sigma} we show
the inclusive $\gamma\gamma+X$ background rates in fb
for different Higgs masses, for $w=1.2\Gamma$ and $w=2\Gamma$
mass window cuts, respectively.
\begin{table*}[ht]
\renewcommand{\arraystretch}{1.5}
\caption{ Background rates in fb for $w=1.2\Gamma$
mass cut, and significance ($S/\sqrt{B}$, for 1 fb$^{-1}$ of data, and
assuming ${\rm B}(h\rightarrow\gamma\gamma)=100\%$) as
a function of the Higgs mass $M_h$. The signal consists of 
associated $Wh/Zh$ production and $WW/ZZ$ fusion. \label{1sigma}}
\begin{tabular}{||c||c||c||c|c|c|c||c|c|c|c||}\hline\hline
      & $\gamma\gamma+X$ 
      & \multicolumn{9}{c||}{Significance $S/\sqrt{B}$}  \\ \cline{3-11}
$M_h$ & bknd &  $\gamma\gamma+X$
      & \multicolumn{4}{c||}{$\gamma\gamma+1$ jet}
      & \multicolumn{4}{c||}{$\gamma\gamma+2$ jets}  \\ \cline{4-11}
(GeV) & (fb) & 
      & $p_T>20$ & $p_T>25$ & $p_T>30$ & $p_T>35$ 
      & $p_T>20$ & $p_T>25$ & $p_T>30$ & $p_T>35$  \\ \hline\hline
100. &271.7 & 16.5 & 31.3 & 34.2 & 36.3 & 35.4 & 31.9 & 36.4 & 35.1 & 31.1 \\ \hline
120. &166.6 & 13.0 & 24.4 & 26.7 & 28.5 & 28.5 & 24.7 & 29.5 & 29.2 & 26.7 \\ \hline
140. &103.0 & 10.7 & 20.1 & 22.4 & 23.9 & 23.9 & 20.6 & 24.0 & 23.7 & 21.0 \\ \hline
160. & 64.3 &  8.9 & 17.0 & 19.1 & 20.1 & 20.4 & 17.2 & 20.2 & 20.8 & 19.2 \\ \hline
180. & 40.5 &  7.3 & 13.5 & 15.0 & 16.2 & 16.5 & 13.6 & 16.3 & 16.7 & 16.2 \\ \hline
200. & 26.1 &  5.8 & 10.6 & 11.9 & 12.5 & 12.7 & 10.4 & 12.3 & 12.7 & 12.0 \\ \hline
250. &  9.4 &  3.7 &  6.7 &  7.5 &  7.9 &  8.2 &  6.6 &  7.9 &  8.7 &  8.4 \\ \hline
300. &  4.8 &  2.2 &  3.8 &  4.3 &  4.7 &  4.7 &  3.6 &  4.2 &  4.9 &  4.7 \\ \hline
350. &  2.3 &  1.5 &  2.8 &  3.1 &  3.3 &  3.3 &  2.4 &  2.8 &  3.0 &  3.4 \\ \hline
400. &  1.2 &  1.0 &  1.8 &  2.0 &  2.0 &  2.1 &  1.7 &  1.7 &  2.1 &  2.4 \\ \hline\hline
\end{tabular}
\end{table*}
\begin{table*}[ht]
\renewcommand{\arraystretch}{1.5}
\caption{ The same as Table \ref{1sigma}, but for a
$w=2\Gamma$ mass window. \label{2sigma}}
\begin{tabular}{||c||c||c||c|c|c|c||c|c|c|c||}\hline\hline
      & $\gamma\gamma+X$ 
      & \multicolumn{9}{c||}{Significance $S/\sqrt{B}$}  \\ \cline{3-11}
$M_h$ & bknd &  $\gamma\gamma+X$
      & \multicolumn{4}{c||}{$\gamma\gamma+1$ jet}
      & \multicolumn{4}{c||}{$\gamma\gamma+2$ jets}  \\ \cline{4-11}
(GeV) & (fb) & 
      & $p_T>20$ & $p_T>25$ & $p_T>30$ & $p_T>35$ 
      & $p_T>20$ & $p_T>25$ & $p_T>30$ & $p_T>35$  \\ \hline\hline
100. &453.4 & 14.4 & 27.0 & 29.8 & 31.7 & 30.6 & 27.7 & 31.9 & 30.5 & 26.4  \\ \hline
120. &278.1 & 11.3 & 21.3 & 23.5 & 25.1 & 24.9 & 21.9 & 25.5 & 25.1 & 22.5  \\ \hline
140. &171.9 &  9.3 & 17.5 & 19.5 & 21.0 & 21.0 & 17.7 & 20.7 & 21.0 & 18.5  \\ \hline
160. &107.3 &  8.0 & 15.1 & 16.7 & 18.0 & 18.2 & 15.4 & 17.8 & 18.2 & 17.3  \\ \hline
180. & 67.6 &  6.6 & 12.2 & 13.7 & 14.6 & 14.9 & 12.4 & 14.3 & 15.2 & 14.2  \\ \hline
200. & 43.6 &  5.4 & 10.1 & 11.4 & 12.1 & 12.1 &  9.9 & 11.9 & 12.3 & 11.1  \\ \hline
250. & 15.7 &  3.6 &  6.5 &  7.3 &  7.7 &  7.8 &  6.4 &  7.6 &  8.5 &  8.0  \\ \hline
300. &  8.1 &  2.1 &  3.8 &  4.2 &  4.6 &  4.5 &  3.6 &  4.3 &  4.5 &  4.5  \\ \hline
350. &  3.9 &  1.6 &  2.7 &  3.0 &  3.4 &  3.4 &  2.6 &  3.1 &  3.5 &  3.1  \\ \hline
400. &  2.1 &  1.1 &  1.9 &  2.2 &  2.3 &  2.4 &  1.6 &  2.1 &  2.6 &  2.4  \\ \hline\hline
\end{tabular}
\end{table*}
Here we have added the intrinsic width $\Gamma_h$ and the experimental
resolution $\Gamma_{\rm exp}=2\sqrt{\ln4}\times\sigma_{\rm exp}\simeq
2.35\times0.15\sqrt{2}\sqrt{E(\gamma)} \simeq 0.35\sqrt{M_h}$
in quadrature: $\Gamma=\left(\Gamma^2_h+\Gamma^2_{\rm exp}\right)^{1/2}$.
The width $\Gamma$ varies between 3.5 GeV for $M_h=100$ GeV and 29.0
GeV for $M_h=400$ GeV. The two tables also show the significance
(for 1 fb$^{-1}$ of data, and assuming
${\rm B}(h\rightarrow\gamma\gamma)=100\%$)
in the inclusive diphoton channel when only
associated $Wh/Zh$ production and $WW/ZZ\rightarrow h$ fusion
are included in the signal sample. We see that (as can also
be anticipated from Fig.~\ref{fig:significance}) a $w=1.2\Gamma$
cut around the Higgs mass typically gives a better
statistical significance, especially for lighter 
(and therefore more narrow) Higgs bosons.

\subsection*{Exclusive channels: analysis}

\begin{figure}[ht]
\epsfysize=2.5in
\epsffile[-150 205 200 565]{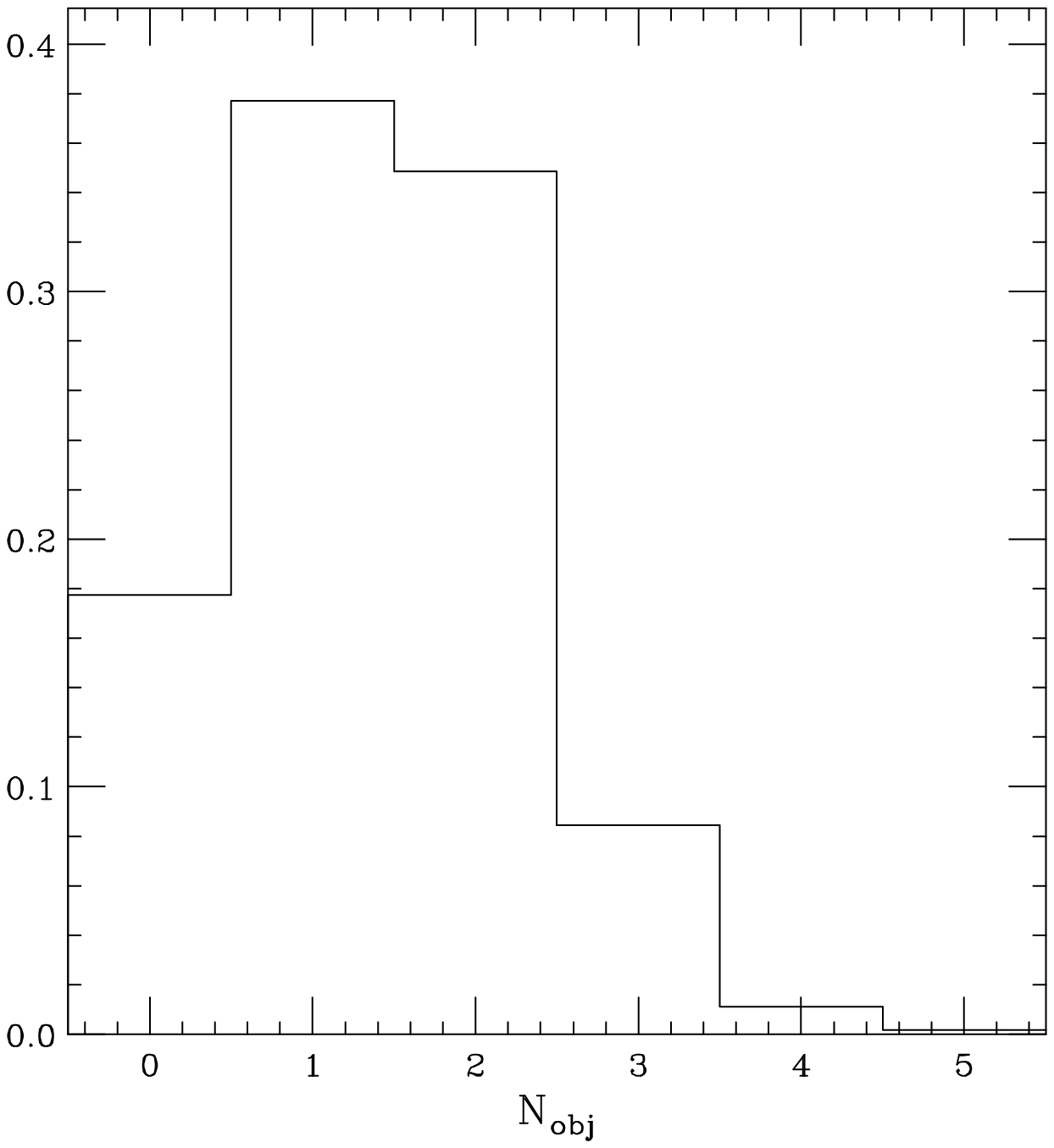}
\begin{center}
\caption[]
{\small The number of ``jets'', which stands for
QCD jets, tau jets and electrons, in associated $Wh$
production, once we require the two photons from the
Higgs to pass the photon ID cuts. \label{nobj}}
\end{center}
\end{figure}

The next question is whether the sensitivity can be further
improved by requiring additional objects in the event.
The point is that a significant fraction
of the signal events from both associated $Wh/Zh$
production and $WW/ZZ$ fusion will have additional hard objects,
most often QCD jets.
In Fig.~\ref{nobj} we show the ``jet'' multiplicity
in associated $Wh$ production, where for detector simulation
we have used the SHW package \cite{SHW} with
a few modifications as in \cite{SHWmod}.
Here we treat ``jets'' in a broader context,
including electrons and tau jets as well.
\begin{figure}[t]
\epsfysize=3.5in
\epsffile[-250 0 100 560]{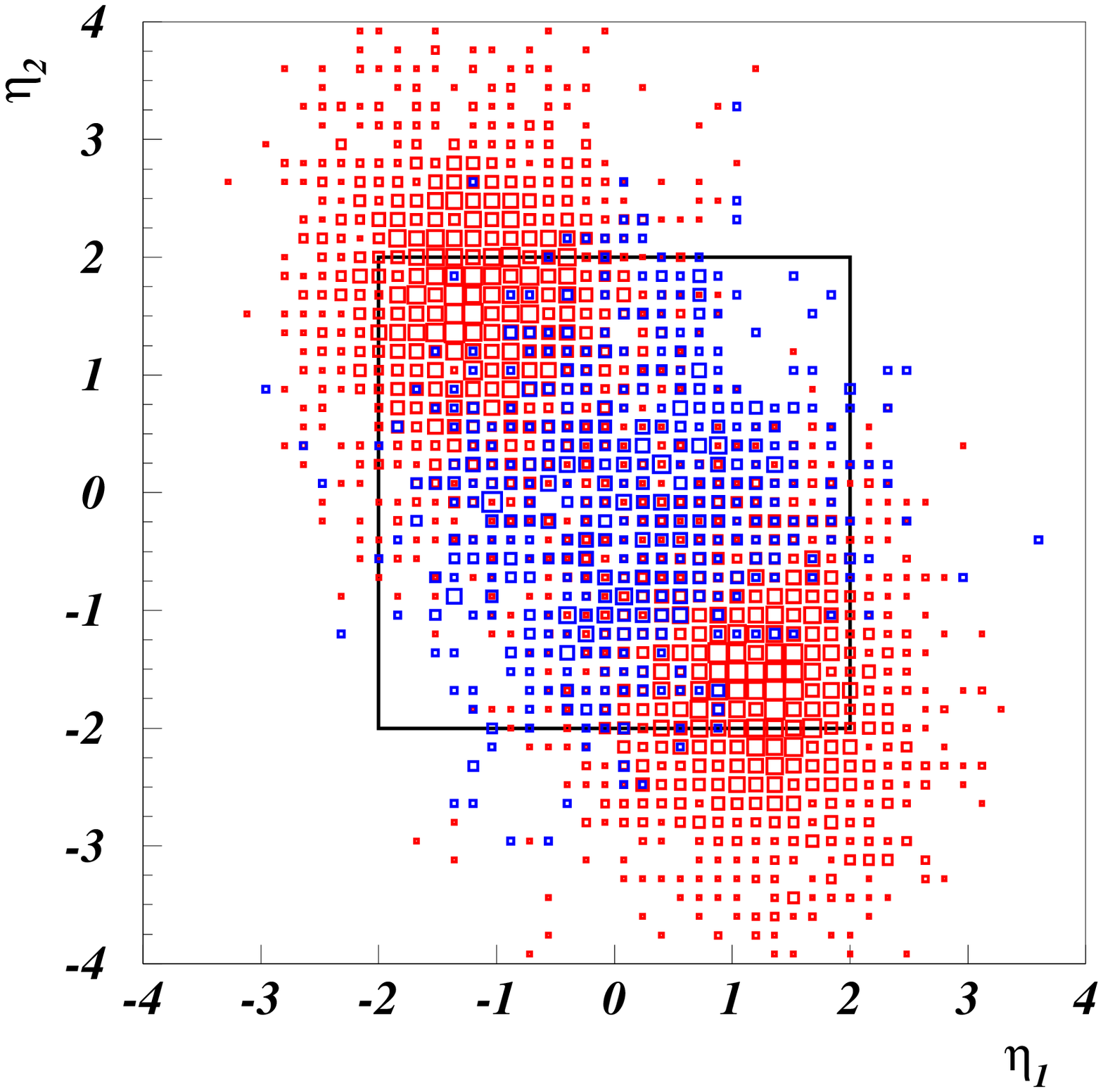}
\begin{center}
\caption[]
{\small Pseudorapidity distribution of the two spectator jets
in $WW/ZZ\rightarrow h$ fusion (red) and 
associated $Wh/Zh$ production (blue). The boxed region
represents the off-line selection cuts.
\label{rapidity}}
\end{center}
\end{figure}

Previous studies \cite{Wilson,D0} have required {\em two}
or more additional QCD jets. Here we shall also consider the
signature with at least {\em one} additional ``jet'',
where a ``jet'' is an object with $|\eta|<2$.
The advantages of not requiring a second ``jet''
are twofold. First, in this way we can also pick up signal from
$WW/ZZ\rightarrow h$ fusion, whose cross-section
does not fall off as steeply with $M_h$, and in fact
for $M_h>200$ GeV is larger than the cross-section for
associated $Wh/Zh$ production\footnote{In the case of a topcolor
Higgs (see the next section) we would also pick up
events with initial state gluon radiation, comprising
about 30\% of the gluon fusion signal, which is the
dominant production process for any Higgs mass.}.
Events from $WW/ZZ\rightarrow h$ fusion typically contain two
very hard forward jets, one of which may easily
pass the jet selection cuts. In Fig.~\ref{rapidity} we show the
pseudorapidity distribution of the two spectator jets
in $WW/ZZ\rightarrow h$ fusion (red) and 
associated $Wh/Zh$ production (blue).
Second, by requiring only one additional jet,
we win in signal acceptance. In order to compensate
for the corresponding background increase,
we shall consider several $p_T$ thresholds for
the additional jet, and choose the one giving the
largest significance.

For the exclusive channels we need to rescale the background
from Fig.~\ref{qcd_background} as follows.
From Monte Carlo we obtain reduction factors of
$4.6\pm0.5$, $6.2\pm1.0$, $7.6\pm1.4$, and $8.6\pm1.5$
for the $\gamma\gamma+1$ jet channel, with $p_T(j)>20$, 25, 30
and 35 GeV, respectively. For the $\gamma\gamma+2$ jets channel
the corresponding background reduction is
$21\pm5$, $38\pm12$, $58\pm21$, and $74\pm26$,
depending on the jet $p_T$ cuts. These scaling factors agree well
with those from the CDF and D\O\ data from Run I.

Notice that we choose not to impose an invariant
dijet mass ($M_{jj}$) cut for the $\gamma\gamma+2$ jets channel.
We do not expect that it would lead to a gain in
significance for several reasons. First, given
the relatively high jet $p_T$ cuts needed for the
background suppression, there will be hardly any
background events left with dijet invariant masses
below the (very wide) $W/Z$ mass window.
Second, the signal events from $WW/ZZ$ fusion,
which typically comprise about $25-30\%$ of our signal,
will have a dijet invariant mass distribution very
similar to that of the background. Finally, not imposing the
$M_{jj}$ cut allows for a higher signal acceptance
because of the inevitable combinatorial ambiguity
for the events with $>2$ jets.

The significances for the two exclusive channels,
with the four different jet $p_T$ cuts,
are also shown in Tables \ref{1sigma} and \ref{2sigma}.
We see that the exclusive $\gamma\gamma+2$ jets channel
with $p_T(j)>30$ GeV typically gives the largest
significance, but our new exclusive $\gamma\gamma+1$ jet
channel is following very close behind.

\begin{figure}[t]
\epsfysize=2.5in
\epsffile[-150 205 200 565]{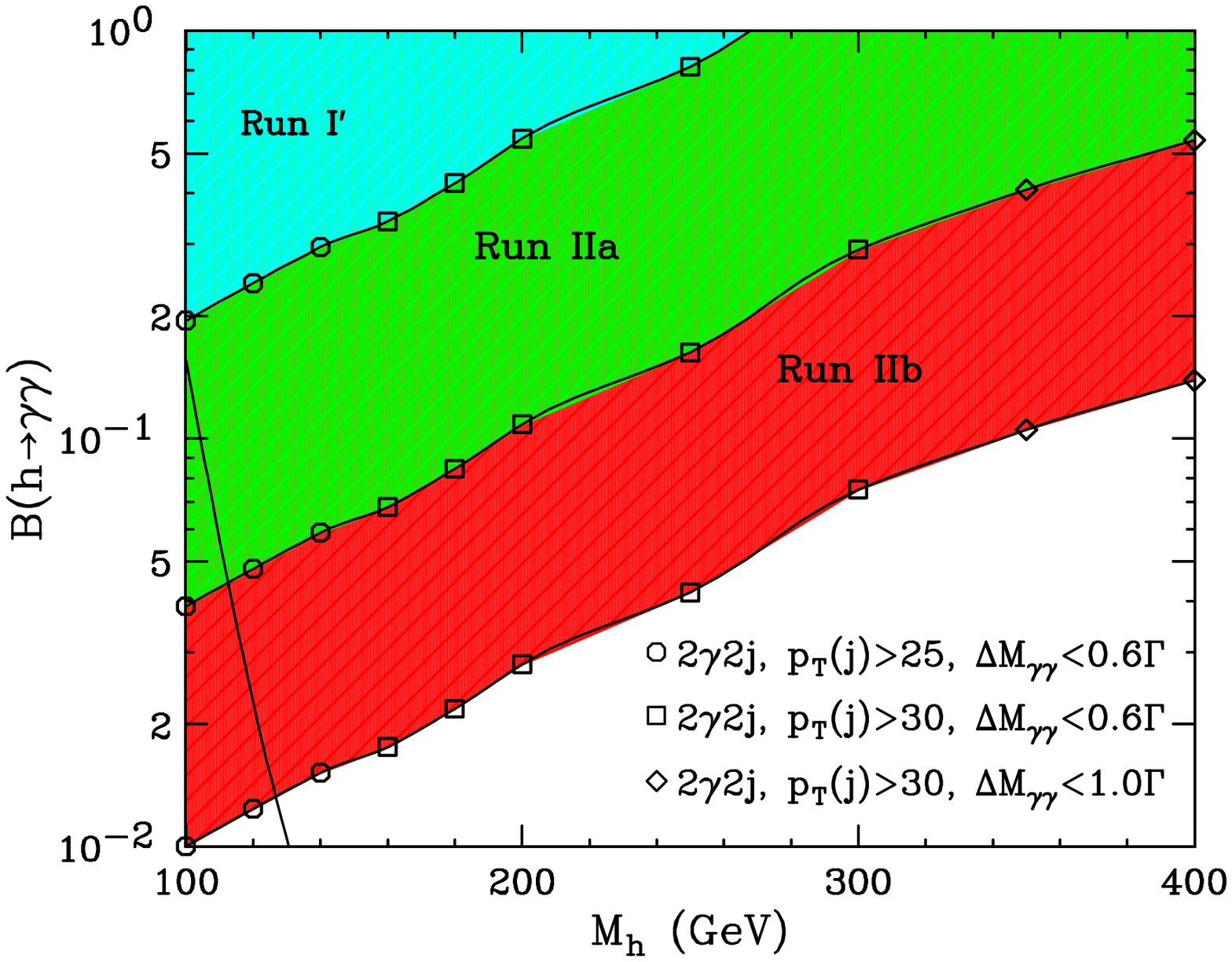}
\begin{center}
\caption[]
{\small 95\% CL upper limit on the branching ratio
${\rm B}(h\rightarrow \gamma\gamma)$, 
with 0.1 (cyan), 2.0 (green) and 30 ${\rm fb}^{-1}$ (red), 
as a function of $M_h$. For each mass point, we compare the
significance for both the inclusive and the exclusive
channels with different cuts, and for the limit we choose the
set of cuts which provides the best reach -
o: $2\gamma+2j$, with $p_T(j)>25$ GeV and $w=1.2\,\Gamma$;
$\Box$: $2\gamma+2j$, with $p_T(j)>30$ GeV and $w=1.2\,\Gamma$; and
$\Diamond$: $2\gamma+2j$, with $p_T(j)>30$ GeV and $w=2.0\,\Gamma$.
The solid line is the prediction for 
the branching ratio of a ``bosonic'' Higgs.
\label{bos}}
\end{center}
\end{figure}

\subsection*{Exclusive channels: results}

We are now ready to present our results for the Run II
Tevatron reach for a bosonic Higgs. In Fig.~\ref{bos} we show
the 95\% CL upper limit on the branching ratio
${\rm B}(h\rightarrow \gamma\gamma)$, 
with 0.1 (cyan), 2.0 (green) and 30 ${\rm fb}^{-1}$ (red), 
as a function of $M_h$. For each mass point, we compare the
significance for both the inclusive as well as the exclusive
channels with all the different cuts, and for the limit
we choose the channel with the set of cuts providing
the best reach. It turns out that for the
case at hand the winners are:
o: $2\gamma+2j$, with $p_T(j)>25$ GeV;
$\Box$: $2\gamma+2j$, with $p_T(j)>30$ GeV, and
$\Diamond$: $2\gamma+1j$, with $p_T(j)>30$ GeV.
In the figure we also show the HDECAY \cite{hdecay}
prediction for ${\rm B}(h\rightarrow \gamma\gamma)$ in case
of a ``bosonic'' Higgs.
The reach shown for 0.1 ${\rm fb}^{-1}$ is intended as a comparison 
to Run I, in fact for the 0.1 ${\rm fb}^{-1}$ curve
we scaled down both the signal and background cross-sections
to their values at 1.8 TeV center-of-mass energy, keeping
the efficiencies the same. In other words, the region marked as
Run I' would have been the hypothetical reach in Run I,
if the improved Run II detectors were available at that time.
As seen from Fig.~\ref{bos}, the reach for a ``bosonic'' Higgs
\cite{bosmodels} (at 95\% CL) in Run IIa and Run IIb is $\sim115$ GeV
and $\sim125$ GeV, correspondingly. This is a significant improvement over
the ultimate reach from LEP \cite{LEP limits} of $\sim 105$ GeV.

\section*{Tevatron Reach for a Topcolor Higgs}

Here we consider the case of a ``topcolor'' bosonic 
Higgs, where the Higgs also couples to
the top and other heavy quarks \cite{topmodels}.
We therefore include events from gluon fusion into our
signal sample. We used the next-to-leading order
cross-sections for gluon fusion from the
HIGLU program \cite{higlu}.

In Tables \ref{1sigmatop} and \ref{2sigmatop}
we show the significance (for 1 fb$^{-1}$ of data, and again
assuming ${\rm B}(h\rightarrow\gamma\gamma)=100\%$)
in the inclusive and the two exclusive channels, for the topcolor Higgs case.
Since gluon fusion, which rarely has additional hard jets,
is the dominant production process,
the inclusive channel typically provides the best reach.
However, the $2\gamma+1j$ channel is again very competitive,
since the additional hard jet requirement manages to suppress
the background at a reasonable signal cost.
We see that our new $2\gamma+1j$ channel clearly
gives a better reach than the $2\gamma+2j$
channel \cite{Lauer,D0,Wilson}. For Higgs masses above $\sim180$ GeV,
it sometimes becomes marginally better even than the 
inclusive diphoton channel. The specific jet $p_T$ cut
and mass window size $w$ seem to be less of an issue --
from Tables~\ref{1sigmatop} and \ref{2sigmatop}
we see that $p_T(j)>25$, $p_T(j)>30$ GeV and $p_T(j)>35$ GeV
work almost equally well, and for $M_h\gsim 200$ GeV
both values of $w$ are acceptable.
\begin{table*}[t]
\renewcommand{\arraystretch}{1.5}
\caption{ The same as Table \ref{1sigma}, but for a
topcolor Higgs, i.e. gluon fusion events are included in the
signal. \label{1sigmatop}}
\begin{tabular}{||c||c||c||c|c|c|c||c|c|c|c||}\hline\hline
      & $\gamma\gamma+X$ 
      & \multicolumn{9}{c||}{Significance $S/\sqrt{B}$}  \\ \cline{3-11}
$M_h$ & bknd &  $\gamma\gamma+X$
      & \multicolumn{4}{c||}{$\gamma\gamma+1$ jet}
      & \multicolumn{4}{c||}{$\gamma\gamma+2$ jets}  \\ \cline{4-11}
(GeV) & (fb) & 
      & $p_T>20$ & $p_T>25$ & $p_T>30$ & $p_T>35$ 
      & $p_T>20$ & $p_T>25$ & $p_T>30$ & $p_T>35$  \\ \hline\hline
100. &271.7 & 54.3 & 43.4 & 44.6 & 45.1 & 42.5 & 33.6 & 37.8 & 36.4 & 32.1\\ \hline
120. &166.6 & 42.5 & 35.3 & 36.9 & 37.3 & 35.3 & 26.5 & 30.8 & 30.2 & 27.5\\ \hline
140. &103.0 & 35.5 & 30.4 & 32.1 & 33.0 & 31.6 & 22.3 & 25.6 & 25.1 & 22.0\\ \hline
160. & 64.3 & 30.1 & 27.3 & 28.8 & 29.2 & 28.3 & 19.5 & 22.4 & 22.4 & 20.3\\ \hline
180. & 40.5 & 24.5 & 22.4 & 23.8 & 24.5 & 23.8 & 15.7 & 18.2 & 18.3 & 17.6\\ \hline
200. & 26.1 & 19.9 & 18.6 & 19.9 & 20.3 & 20.0 & 12.7 & 14.4 & 14.3 & 13.3\\ \hline
250. &  9.4 & 12.9 & 12.5 & 13.3 & 13.6 & 13.4 &  8.4 &  9.7 & 10.2 &  9.8\\ \hline
300. &  4.8 &  9.1 &  8.7 &  9.3 &  9.6 &  9.4 &  4.9 &  5.6 &  5.9 &  5.9\\ \hline
350. &  2.3 &  8.4 &  7.9 &  8.4 &  8.4 &  7.9 &  4.0 &  4.1 &  4.5 &  4.5\\ \hline
400. &  1.2 &  6.4 &  6.1 &  6.1 &  6.3 &  6.2 &  3.0 &  3.0 &  3.5 &  3.1\\ \hline\hline
\end{tabular}
\end{table*}
\begin{table*}[t]
\renewcommand{\arraystretch}{1.5}
\caption{ The same as Table \ref{1sigmatop}, but for a
$w=2\Gamma$ mass window. \label{2sigmatop}}
\begin{tabular}{||c||c||c||c|c|c|c||c|c|c|c||}\hline\hline
      & $\gamma\gamma+X$ 
      & \multicolumn{9}{c||}{Significance $S/\sqrt{B}$}  \\ \cline{3-11}
$M_h$ & bknd &  $\gamma\gamma+X$
      & \multicolumn{4}{c||}{$\gamma\gamma+1$ jet}
      & \multicolumn{4}{c||}{$\gamma\gamma+2$ jets}  \\ \cline{4-11}
(GeV) & (fb) & 
      & $p_T>20$ & $p_T>25$ & $p_T>30$ & $p_T>35$
      & $p_T>20$ & $p_T>25$ & $p_T>30$ & $p_T>35$ \\ \hline\hline
100. &453.4 & 47.7 & 37.6 & 38.9 & 39.5 & 36.9 & 29.0 & 33.0 & 31.6 & 27.1\\ \hline
120. &278.1 & 37.7 & 31.2 & 32.6 & 33.1 & 31.2 & 23.5 & 26.6 & 25.9 & 23.1\\ \hline
140. &171.9 & 31.4 & 26.7 & 28.2 & 29.0 & 27.9 & 19.2 & 22.1 & 22.0 & 19.4\\ \hline
160. &107.3 & 26.6 & 24.3 & 25.3 & 26.0 & 25.2 & 17.4 & 19.7 & 19.5 & 18.3\\ \hline
180. & 67.6 & 22.3 & 20.5 & 21.7 & 22.1 & 21.6 & 14.2 & 16.2 & 16.8 & 15.4\\ \hline
200. & 43.6 & 18.8 & 17.7 & 19.1 & 19.4 & 18.9 & 12.0 & 13.7 & 13.8 & 12.5\\ \hline
250. & 15.7 & 12.8 & 12.1 & 13.2 & 13.4 & 13.1 &  8.2 &  9.3 & 10.0 &  9.6\\ \hline
300. &  8.1 &  9.0 &  8.5 &  9.1 &  9.2 &  9.0 &  5.0 &  5.6 &  5.6 &  5.7\\ \hline
350. &  3.9 &  8.3 &  7.6 &  8.1 &  8.1 &  7.9 &  4.0 &  4.7 &  4.6 &  4.4\\ \hline
400. &  2.1 &  6.3 &  6.1 &  6.3 &  6.3 &  6.1 &  2.9 &  3.4 &  3.7 &  3.6\\ \hline\hline
\end{tabular}
\end{table*}
\begin{figure}[t]
\epsfysize=2.5in
\epsffile[-150 205 200 565]{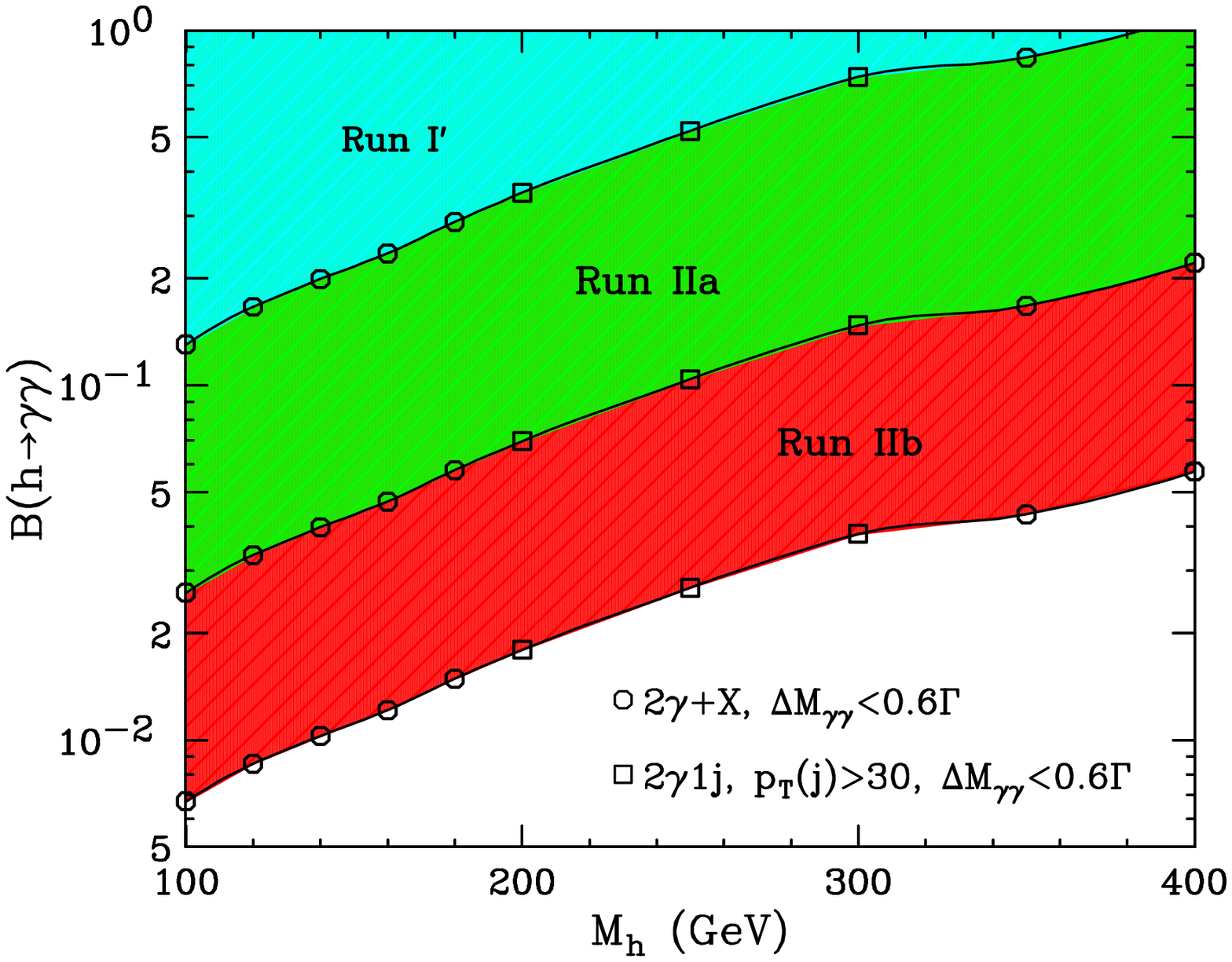}
\begin{center}
\caption[]
{\small The same as Fig.~\ref{bos}, but for
a topcolor Higgs, i.e. gluon fusion events are
included in the signal. The channels with the best
$S/\sqrt{B}$ ratio are: 
o: inclusive $2\gamma+X$, and 
$\Box$: $2\gamma+1j$, with $p_T(j)>30$ GeV;
both with $w=1.2\,\Gamma$.
\label{topcolor}}
\end{center}
\end{figure}
In Fig.~\ref{topcolor} we show the Run II reach for the
branching ratio ${\rm B}(h\rightarrow \gamma\gamma)$
as a function of the Higgs mass, for the case
of a ``topcolor'' Higgs boson. This time the channels
with the best signal-to-noise ratio are:
o: inclusive $2\gamma+X$, and
$\Box$: $2\gamma+1j$, with $p_T(j)>30$ GeV;
both with $w=1.2\Gamma$.

\section*{Conclusions}

We have studied the Tevatron reach for Higgs bosons decaying into
photon pairs. For purely ``bosonic'' Higgses, which only couple to
gauge bosons, the $2\gamma+2j$ channel offers the best reach, but
the $2\gamma+1j$ channel is almost as good.
For topcolor Higgs bosons, which can also be produced via
gluon fusion, the inclusive $2\gamma+X$ channel is the best,
but the $2\gamma+1j$ channel is again very competitive.
We see that in both cases the $2\gamma+1j$ channel is a no-lose
option!

\vspace*{0.5cm}
{\bf Acknowledgments.}
We would like to thank S.~Mrenna for many useful discussions
and B.~Dobrescu for comments on the manuscript.
This research was supported in part by the U.S.~Department of Energy
under Grants No.~DE-AC02-76CH03000 and DE-FG02-91ER40688.
Fermilab is operated under DOE contract DE-AC02-76CH03000.

\end{document}